\documentclass[twocolumn,showpacs,preprintnumbers,amsmath,amssymb]{revtex4}
\usepackage{graphicx}
\usepackage{dcolumn}
\usepackage{bm}
\begin{document}
\title{ Neutron stars, $\beta$-stable ring-diagram equation of state \\
and Brown-Rho scaling}
\author{Huan\ Dong and T.\ T.\ S.\ Kuo }
\affiliation{Department of Physics and Astronomy,
Stony Brook University, Stony Brook,NY 11794-3800, USA}
\email{thomas.kuo@stonybrook.edu}
\author{R.\ Machleidt  }
\affiliation{Department of Physics,
University of Idaho, Moscow, Idaho 83844, USA}

\begin{abstract}
Neutron star properties, such as its
mass, radius, and moment of inertia, are calculated by solving the 
Tolman-Oppenheimer-Volkov ({\rm TOV}) equations using the ring-diagram
equation of state ({\rm EOS}) obtained from realistic low-momentum {\rm NN}
interactions $V_{low-k}$. Several {\rm NN} potentials  (CDBonn,
Nijmegen, Argonne V18 and BonnA) have been employed to
 calculate the ring-diagram {\rm EOS}
where the particle-particle hole-hole ring diagrams are summed to all orders. 
The proton fractions for different radial regions
of a $\beta$-stable neutron star are determined from the chemical potential
 conditions $\mu_n-\mu_p=\mu_e=\mu_{\mu}$. 
The neutron star masses, radii and moments of inertia 
given by the above potentials all tend to be too small compared with
the accepted values. Our results are largely improved with the inclusion
of  a Skyrme-type three-body force  based on Brown-Rho scalings where 
the in-medium meson masses, particularly those
of $\omega$, $\rho$ and $\sigma$, are slightly decreased compared with
their in-vacuum values.  Representative results using such medium corrected
 interactions are maximum neutron-star mass \textit{M}$\sim 1.8 M_{\odot}$  
with radius \textit{R}$\sim 9$ km and moment of inertia $\sim 60 M_{\odot}km^2$,
values given by the four {\rm NN} potentials  being nearly the same.
The effects of nuclei-crust {\rm EOS}s on properties of neutron stars
 are discussed. 
\end{abstract}
\pacs{pacs} \maketitle

\section{Introduction}
 Neutron stars are  very interesting physical systems   
and their properties, such as masses and radii, 
can be derived from the equation of state ({\rm EOS}) of the nuclear medium
contained in them. In  carrying out such derivation,
there is, however, a well-known difficulty, namely the {\rm EOS} is not fully
known. Determination of the {\rm EOS} for neutron stars is an important yet
challenging undertaking. As reviewed in  
\cite{latti07,prakash01,sedra07,weber99,shapiro},
this topic has been extensively studied and much progress has been made.
Generally speaking, there are two complementary approaches to determine
the {\rm EOS}. One is to deduce it from heavy-ion collision experiments,
and crucial information about the {\rm EOS} has already been obtained 
\cite{daniel02,latti07,li08,tsang09}.
Another approach is to calculate the {\rm EOS} microscopically
from a many-body theory. (See, e.g. \cite{prakash01,heisel00,sam09} and 
references quoted therein.)
As is well-known, there are a number of difficulties
in this approach. Before discussing them,
 let us first briefly outline the derivation of
neutron-star properties from its {\rm EOS}. One starts
from the Tolman-Oppenheimer-Volkov ({\rm TOV}) equations
\begin{eqnarray}
\frac{dp(r)}{dr}&=&-\frac{GM(r)\epsilon(r)}{c^2r^2}
                  \frac{[1+\frac{p(r)}{\epsilon(r)}]
                  [1+\frac{4\pi r^3p(r)}{M(r)c^2}]}  
                  {[1-\frac{2GM(r)}{rc^2}]},  \nonumber \\
\frac{dM(r)}{dr}&=&4\pi r^2\epsilon(r) 
\end{eqnarray}
where \textit{p}(\textit{r}) is the pressure at radius \textit{r} and \textit{M}(\textit{r}) is the gravitational mass
inside \textit{r}. \textit{G} is the gravitational constant 
and $\epsilon (r)$ is the energy 
density inclusive of the rest mass density. The solutions of these equations
are obtained by integrating them out from the neutron-star center
till its edge where \textit{p} is zero. (Excellent pedagogical reviews on 
neutron stars and {\rm TOV} equations can be found in e.g. \cite{silbar,shapiro}.) 
 In solving the above equations, 
an indispensable ingredient is clearly the nuclear matter {\rm EOS} for 
the energy density $\epsilon (n)$, \textit{n} being the medium density. 
As the density at the neutron star center is typically very high
(several times higher than normal nuclear
saturation density of $n_0\simeq 0.16 fm^{-3}$), we need to have  
the above {\rm EOS} over a wide range of densities, from very low to very high.

In the present work we shall calculate the nuclear {\rm EOS} directly from
a fundamental nucleon-nucleon ({\rm NN}) interaction $V_{NN}$ and then 
use it to calculate neutron star 
properties by way of the {\rm TOV} equations. 
There have been  neutron-star calculations using a number of {\rm EOS}s, most
of which  empirically determined, and the mass-radius trajectories given
by them are widely different from each other (see, e.g. Fig. 2 of 
\cite{latti07}). To determine the {\rm EOS} with less
uncertainty would certainly be desirable.
 There are a number of different {\rm NN} potential models such as  the CDBonn 
\cite{cdbonn}, Nijmegen \cite{nijmegen}, Argonne V18 
\cite{argonne} and BonnA \cite{mach89} potentials. 
 These potentials all possess strong short-range repulsions
and to use them in many-body calculations one needs first take care of
their short-range correlations by way of some renormalization methods.
 We shall use in the present work the recently developed renormalization
group  method which converts $V_{NN}$ into an effective 
low-momentum {\rm NN} interaction $V_{low-k}$
\cite{bogner01,bogner02,coraggio02,schwenk02,bogner03,jdholt}.
An advantage of this interaction is its near uniqueness, in the sense that
the $V_{low-k}$s derived from different realistic {\rm NN} interactions
are nearly the same. Also $V_{low-k}$ is a smooth potential
suitable for being directly used in many-body calculations.
This $V_{low-k}$ will then be used to calculate the nuclear {\rm EOS} using
 a recently developed low-momentum ring-diagram
approach \cite{siu09}, where the particle-particle hole-hole (\textit{pphh})
ring diagrams of the {\rm EOS} are summed  to all orders.
The above procedures will be discussed in more detail later on.

 We shall also study
the effects of Brown-Rho ({\rm BR}) scaling \cite{brown91,brown96,brown02,brown04} 
on neutron star
properties. As discussed in \cite{siu09}, low-momentum ring diagram
calculations using two-body {\rm NN} interactions alone are not able to reproduce
the empirical properties for symmetric nuclear matter; the calculated
 energy per particle ($E_0/A$) and saturation density ($n_0$)
 are both too high compared with the empirical values
of $E_0/A\simeq$-16 {\rm MeV} and $n_0\simeq0.16~fm^{-3}$. A main idea of the
{\rm BR} scaling is that the masses of in-medium mesons are generally
 suppressed, because of their interactions with the background medium,
 compared with   mesons in free space. 
As a result, the {\rm NN} interaction in the nuclear medium can be 
significantly different
from that in free space, particularly at high density. Effects from such
medium modifications have been found to be very helpful in reproducing
the empirical properties of symmetric nuclear matter \cite{siu09}.
 Dirac-Brueckner-Hartree-Fock ({\rm DBHF}) nuclear matter calculations
have been conducted with and without {\rm BR} 
scaling~\cite{rapp99,alonso,brown87}.  
In addition, {\rm BR} scaling has played an essential role in explaining
the extremely long life time of $^{14}C$ $\beta$-decay \cite{holt08}. 
As  mentioned earlier, the central density of neutron stars
is typically rather high, $\sim 8n_0$ or higher,  $n_0$ being the 
saturation density of normal nuclear matter. At such high density, 
the effect of {\rm BR} scaling
should be especially significant. Neutron stars may provide an important
 test for {\rm BR} scaling.  

 In the following, we shall first  describe in section II the  
derivation of the low-momentum {\rm NN} interaction $V_{low-k}$, on which our
ring-diagram {\rm EOS} will be based.  Our method for the
all order summation of the ring diagrams   will also be addressed. 
Previously such summation has been carried out for neutron matter \cite{siu08}
and for symmetric nuclear matter \cite{siu09}. In the present work
we consider  $\beta$-stable nuclear matter composed of neutrons, protons,
electrons, and muons. Thus we need to calculate
ring diagrams for asymmetric nuclear matter whose neutron and proton
fractions are different. An improved treatment for the angle averaged
proton-neutron Pauli exclusion operator will be discussed.
In section III we shall outline the {\rm BR} scaling for in-medium {\rm NN} interactions.
We shall discuss that the effect of {\rm BR} scaling can be satisfactorily
simulated by an empirical three-body force of the Skyrme type.
 In section IV, we shall present and discuss our results of neutron star 
calculations based on ring-diagram pure-neutron {\rm EOS}s, 
ring-diagram $\beta$-stable {\rm EOS}s consisted of
neutrons, protons, electrons
 and muons, and the well-known {\rm EOS} of Baym, Pethick and 
Sutherland ({\rm BPS})
\cite{baym71} for the nuclei crust of neutron stars.

\section{ Ring-diagram {\rm EOS} for asymmetric nuclear matter}
 In our calculation of neutron star properties, we shall employ a 
nuclear {\rm EOS}
derived microscopically from realistic {\rm NN} potentials
$V_{NN}$. Such microscopic calculations
would provide a test if it is possible
to derive neutron star properties starting from an underlying {\rm NN} 
interaction.
In this section, we shall describe the methods for this derivation.
A first step in this regard is to derive  an effective low-momentum
interaction $V_{low-k}$ by way of a renormalization procedure,
the details of which have been described in
\cite{bogner01,bogner02,coraggio02,schwenk02,bogner03,jdholt}.
Here we shall just briefly outline its main steps. 
It is generally believed that  the low-energy properties of physical
systems can be satisfactorily described by an effective theory
confined within a low-energy (or low-momentum) model space \cite{bogner03}.
In addition, the high-momentum (short range) parts of various $V_{NN}$
models are model dependent and rather uncertain \cite{bogner03}.

Motivated by the above two considerations, the following low-momentum
renormalization (or model-space) approach has been introduced. Namely  
one employes a low-momentum model space where all particles have
momentum less than a cut-off scale  $\Lambda$. The corresponding 
renormalized effective {\rm NN} interaction is $V_{low-k}$, which is obtained
by integrating out the $k>\Lambda$ momentum components of $V_{NN}$.
This ``integrating-out" procedure is carried out by way of a \textit{T}-matrix 
equivalence
approach. We start from the full-space
 \textit{T}-matrix equation
\begin{multline}
  T(k',k,k^2) = V_{NN}(k',k) \\
 + {\cal P} \int _0 ^{\infty} q^2 dq  \frac{V_{NN}(k',q)T(q,k,k^2 )} {k^2-q^2},  
\label{vlk1}
\end{multline}
where $\cal P$ denotes the principal-value integration.
Notice that in the above equation the intermediate state momentum $q$ is
integrated from 0 to $\infty$.
We then define an effective low-momentum $T$-matrix by
\begin{multline}
T_{low-k }(p',p,p^2) = V_{low-k }(p',p) \\
 + {\cal P} \int_0 ^{\Lambda} q^2 dq \frac{V_{low-k }(p',q)T_{low-k}
 (q,p,p^2)} {p^2-q^2 },
\label{vlk2}
\end{multline}
where the intermediate state momentum is integrated from
0 to $\Lambda$, the momentum space cutoff. The low momentum interaction
$V_{low-k}$ is then obtained from the above equations by  requiring the
$T$-matrix equivalence condition to hold, namely
\begin{equation}
 T(p',p,p^2 ) = T_{low-k }(p',p, p^2 ) ;~( p',p) \leq \Lambda.
\label{vlk3}
\end{equation}
Note that $T$ and $T_{low-k}$ are both half on energy shell, and they are equivalent within $\Lambda$. The low-energy ($<\Lambda ^2$) phase shifts of
$V_{NN}$ are preserved by $V_{low-k}$ and so is the deuteron binding energy.
As we shall discuss later, for neutron star calculations we need to
employ a cut-off scale of $\Lambda \sim 3~ fm^{-1}$.

For many years, the Brueckner-Hartree-Fock ({\rm BHF})  
\cite{bethe,mach89,baldo,baldo2,jwholt} and the {\rm DBHF} 
\cite{brockmann} methods were
the primary framework for nuclear matter calculations. ({\rm DBHF} is a relativistic
generalization of {\rm BHF}). In both {\rm BHF} and {\rm DBHF} the  G-matrix interaction is 
employed. This G-matrix interaction is energy dependent. This energy
dependence adds complications to calculations. 
In contrast, the above $V_{low-k}$ is energy independent which facilitates
the calculation of nuclear {\rm EOS}.

 We shall use $V_{low-k}$ to calculate the {\rm EOS} of asymmetric nuclear
matter of total density $n$ and asymmetric parameter $\alpha$
\begin{equation}
 n=n_n +n_p; ~~\alpha=\frac{n_n-n_p}{n_n+n_p},
\end{equation}
where $n_n$ and $n_p$ denote respectively the neutron and proton density
and they are related to the respective Fermi momentum by 
$k_{Fn}^3/(3\pi^2)$ and
$k_{Fp}^3/(3\pi^2)$. The proton fraction is $\chi=(1-\alpha)/2$.

\begin{figure}[here]    
\scalebox{0.4}{\includegraphics{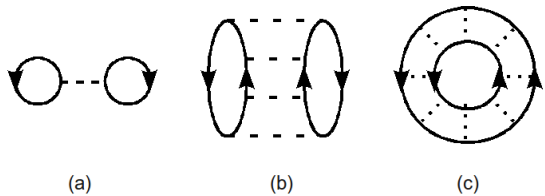}}
\caption{Diagrams included in the all-order \textit{pphh} ring-diagram 
summation for the ground state energy shift of nuclear matter.
Each dashed line represents a $V_{low-k}$ vertex.}
\end{figure}

In our ring-diagram {\rm EOS} calculation, the ground-state energy shift $\Delta E$ 
is given by the all-order sum of the \textit{pphh} ring diagrams as illustrated
in Fig. 1, where (a), (b) and (c) are respectively 1st-, 4th- and 8th-order
such diagrams. ($\Delta E_0$ is defined as ($E_0-E_0^{free}$) where $E_0$
is the  true ground state energy and $E_0^{free}$ that for the non-interacting
system.) Following \cite{siu09}, we have
\begin{multline}\label{eng}
\Delta E_0(n,\alpha)=\int_0^1 d\lambda
\sum_m \sum_{ijkl<\Lambda}Y_m(ij,\lambda) \\ \times Y_m^*(kl,\lambda) \langle
ij|V_{low-k}|kl \rangle,
\end{multline}
where the transition amplitudes are given by the {\rm RPA} equation
 \begin{multline}
\sum _{kl}[(\epsilon_i+\epsilon_j)\delta_{ij,kl}+
\lambda(\bar f_i\bar f_j -f_if_j)\langle ij|V_{low-k}|kl\rangle] \\
\times Y_n(kl,\lambda)
=\omega_nY_n(ij,\lambda);~~(i,j,k,l)\leq\Lambda. \label{rpa}
\end{multline}
In the above, the single particle (s.p.) indices $(i,j,...k,l)$ denote
both protons and neutrons. The s.p. energies $\epsilon$ are the Hartree-Fock
energies given by
\begin{equation}
 \epsilon_k = 
  \hbar^2k^2/2m +\sum_{h < k_F(h)}\langle kh|V_{low-k}|kh\rangle,
\label{sp}
\end{equation}
where $k_F(h)=k_{Fn}$ if $h$ is neutron and $=k_{Fp}$ if it is proton.
Clearly $\epsilon_k$ depends on the nuclear matter density $n$
and the asymmetric parameter $\alpha$. The  occupation
factors $f_i$ and $f_j$ of Eq.(7) are given by
 $f_a=1$ for $k \leq k_F(a)$ and $f_a=0$ for $k>k_F(a)$; also $\bar
f_a=(1-f_a)$.
Again  $k_F(a)=k_{Fn}$ if $a$ is a neutron and $=k_{Fp}$ if it is a proton.
The factor $(\bar f_i \bar f_j -f_if_j)$ is clearly also dependent on $n$
and $\alpha$.
Note that the normalization condition for $Y_m$ in Eq.(6) is
$\langle Y_m|\frac{1}{Q}|Y_m\rangle=-1$ and
$Q(\vec k_i, \vec k_j)=(\bar f_i \bar f_j -f_if_j)$ \cite{tzeng94}. In addition,
$\underset{m}{\Sigma}$ in Eq.(6) means
we sum over only those solutions of the {\rm RPA} equation (\ref{rpa}) which are
dominated by hole-hole components as indicated by the normalization condition.
Note that there is a strength parameter $\lambda$ in the above, and it is 
integrated from 0 to 1. 

 The amplitudes \textit{Y} in Eq.(6) actually represent the overlap
matrix elements
\begin{equation}
Y_m^*(kl,\lambda)=\langle \Psi_m(\lambda,A-2)|a_l
a_k|\Psi_0(\lambda,A)\rangle,
\end{equation}
where $\Psi_0(\lambda,A)$ denotes the true ground state of nuclear matter
which has \textit{A} particles while
 $\Psi_m(\lambda,A-2)$ the \textit{m}th true eigenstate of the \textit{(A-2)} system.
If there is no ground-state correlation (i.e. $\Psi_0$ is a closed Fermi
sea), we have
$Y_m^*(kl,\lambda)=f_kf_l$ and Eq.(6) reduces to the HF result.
Clearly our {\rm EOS} includes effect of ground-state correlation
generated by the all-order sum of \textit{pphh} ring diagrams.

A computational aspect may be mentioned. We shall solve ring-diagram equations
 on the relative and center-of-mass ({\rm RCM}) coordinates
$\vec k$ and $\vec K$ \cite{song87}. 
($\vec k=(\vec k_i -\vec k_j)/2$ and $\vec K=\vec k_i +\vec k_j$.)
In so doing, the treatment of the Pauli operator  
$Q(\vec k_i,\vec k_j)\equiv (\bar f_i\bar f_j-f_if_j)$ of Eq.(7) plays
an important role. This operator is defined in the laboratory frame,  
with its value being  either 1 or -1  (for $pphh$ ring diagrams). 
In our calculation, however, we need \textit{Q} in the
{\rm RCM} coordinates. Angle-average
approximations are commonly used in nuclear matter calculations, and with
them we can obtain angle-averaged $\bar Q(k,K)$. For symmetric nuclear
matter, detailed expressions for $\bar Q$ have been given, see Eqs.(4.9)
and (4.9a) of \cite{song87}. The results for this case, where the Fermi
surfaces for neutron and proton are equal, are already
fairly complicated. 
Derivation of the angle-averaged Pauli operator for asymmetric nuclear
matter has been worked out in \cite{song92}. Their  derivation 
and results  are both considerably more  complicated than the symmetric
case. It would be desirable if they can be simplified. We have found 
that this can be attained by way of a scale transformation. Namely we introduce
new s.p. neutron and proton momentum coordinates 
$\vec k'_n$ and  $\vec k'_p$  defined by 
\begin{equation}
\vec k'_n=\vec k_n \sqrt{k_{Fp}/k_{Fn}};~~  
 \vec k'_p=\vec k_p \sqrt{k_{Fn}/k_{Fp}}.
\end{equation}
 On this new frame, the Fermi surfaces
for neutron and proton become equivalent, both being $(k_{Fn}k_{Fp})^{1/2}$;
 asymmetric nuclear matter can then be treated effectively as a symmetric one
as far as the Pauli operator is concerned.   
We have found that this transformation largely simplifies the derivation and 
calculation of the angle-averaged asymmetric Pauli operator $\bar Q$ and 
will be employed in the present work. 

\section{ In-medium {\rm NN} interactions based on Brown-Rho scaling}
\label{brss}
A main purpose of the present work is to study whether neutron star properties
can be satisfactorily described by {\rm EOS} microscopically derived from
{\rm NN} interactions. Before proceeding, it is important to also check
if such {\rm EOS} can satisfactorily describe the properties of normal 
nuclear matter. The {\rm EOS} and {\rm NN} interaction one uses for neutron star should also be applicable
to normal nuclear matter.
As discussed in  \cite{siu09},  many-body
 calculations for normal nuclear matter using 
two-body {\rm NN} interactions alone are generally not capable in reproducing empirical
 nuclear matter saturation properties. 
To remedy this shortcoming one needs to
consider three-body forces and/or  
{\rm NN} interactions with in-medium modifications \cite{rapp99,bogner05,siu09}.

A central result of the {\rm BR} scaling  is that the masses of mesons in nuclear 
medium are suppressed (dropped) compared to those in 
free space \cite{brown91,brown96,brown02,brown04}.
Nucleon-nucleon interactions are mediated by meson exchange,
and clearly  in-medium modifications of meson masses can significantly
alter the {\rm NN} interaction. These modifications
could arise from the partial restoration of chiral symmetry at finite
density/temperature or from traditional many-body effects.
Particularly important
are the vector mesons, for which there is now evidence from both theory
\cite{hatsuda, harada, klingl} and experiment
\cite{metag, naruki}
that the masses may decrease by approximately $10-15\%$ at normal nuclear
matter density and zero temperature.  For densities below that of 
nuclear matter, a linear approximation for the in-medium mass decrease has been
 suggested \cite{hatsuda}, namely 
\begin{equation}
\frac{m_V^*}{m_V} = 1- C \frac{n}{n_0},
\label{brs}
\end{equation}
where $m_V^*$ is the vector meson mass in-medium, $n$ is the local nuclear
matter density and $n_0$  the nuclear matter
saturation density. \textit{C} is a constant
of value $\sim 0.10-0.15$. {\rm BR} scaling has been found to be very important
for nuclear matter saturation properties in  the ring-diagram calculation
of symmetric nuclear matter of \cite{siu09}.

It is of interest that the effect of {\rm BR} scaling in nuclear matter can
be well represented by an empirical Skyrme three-body force. \cite{siu09}
The Skyrme force has been a widely used effective interaction in 
nuclear physics and it has been very successful in describing the 
properties of both finite nuclei and nuclear matter\cite{skyrme}. It has both
two-body and three-body terms, namely
\begin{equation}
V_{Skyrme}=\sum_{i<j} V(i,j) +\sum_{i<j<k} V_{3b}(i,j,k).
\end{equation}
Here $V(i,j)$ is a momentum  dependent zero-range interaction.
Its three-body term is also a zero-range interaction
\begin{equation}
V_{3b}(i,j,k)=t_3\delta(\vec r_i-\vec r_j)\delta(\vec r_j-\vec r_k)
\end{equation}
which is usually expressed as  a density-dependent two-body interaction
of the form
\begin{equation}
 V_{n}(1,2)=\frac{1}{6}(1+x_3P_{\sigma})t_3 \delta (\vec r_1-\vec r_2)
n (\vec r_{av}),
\end{equation}
 where $P_{\sigma}$ is the spin-exchange operator and
 $\vec r_{av}= \frac{1}{2}(\vec r_1 +\vec r_2)$. $t_3$ and $x_3$ are parameters
determined by fitting certain experimental data.
The general structure of $V_{Skyrme}$ is rather
similar to the effective interactions based on effective field theories
 (EFT) \cite{bogner05},
with $V(i,j)$ corresponding to $V_{low-k}$ and $V(i,j,k)$
to the EFT three-body force. The Skyrme three-body force, however, 
is much simpler than that in EFT.

\section{Results and discussions}
\subsection{Symmetric nuclear matter and {\rm BR} scaling}
  When an {\rm EOS} is used to calculate  neutron-star properties, it is important
and perhaps necessary to first test if the {\rm EOS} can satisfactorily
describe the properties of symmetric nuclear matter such as its 
energy per particle $E_0/A$ and saturation density $n_0$. In principle, only 
those {\rm EOS}s which have done well in this test are suitable for being 
used in neutron star calculation. In this subsection, we shall
calculate properties of symmetric nuclear matter using the
low-momentum ring-diagram {\rm EOS} which we will use in our neutron star
calculations, to test if it can meet the above test.
As described in section II, we first calculate the $V_{low-k}$
interaction for a chosen decimation scale $\Lambda$. Then we calculate
the ground state energy per particle $E_0/A$ using Eq.(6) 
($E_0=E_0^{free}+\Delta E_0.$) with the $pphh$ ring diagrams summed to all
orders. 

In the above calculation, the choice of $\Lambda$ plays an important role.
As discussed in \cite{siu09},
the tensor force is important for nuclear saturation and therefore one should
use a sufficiently large $\Lambda$ so that the tensor force is not integrated out
during the derivation of $V_{low-k}$. Since the main momentum components
of the tensor force has $k\sim 2 fm^{-1}$, one needs to use $\Lambda \sim 3 fm^{-1}$
or larger. 
There is another consideration concerning the choice of $\Lambda$. The density
of neutron star interior is very high, several times larger than $n_0$.
To accommodate such high density, it is necessary to use sufficiently
large $\Lambda$, suggesting  a choice of $\Lambda$ larger than $\sim 3 fm^{-1}$.
As discussed in \cite{bogner03}, a nice
feature of $V_{low-k}$ is its near uniqueness: The $V_{low-k}$s derived from
various different realistic {\rm NN} potentials are practically identical to each
other for $\Lambda < \sim 2.1 fm^{-1}$, while for larger $\Lambda$s the resulting
$V_{low-k}$s begin to have noticeable differences 
but are still similar to each other for $\Lambda$
up to about 3.5$fm^{-1}$. This and the above considerations have led us
to choose $\Lambda$ between $\sim3$ and $\sim3.5fm^{-1}$ for our present
study. The dependence of our results on the choice on $\Lambda$ will be
discussed later on.

\begin{figure}[here]     
\scalebox{0.42}{\includegraphics[angle=-90]{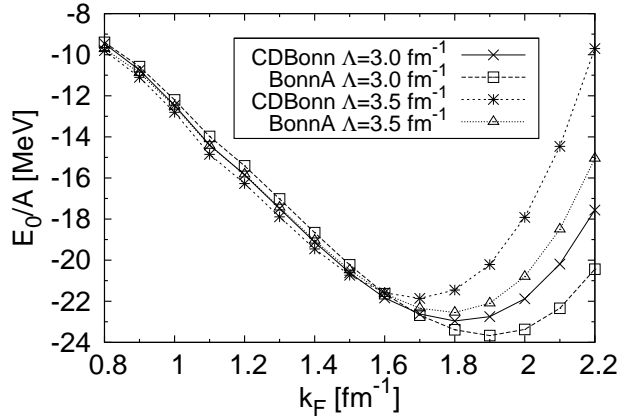}}
\caption{ Ring-diagram {\rm EOS}s for symmetric nuclear matter 
with $V_{low-k}$ derived from CDBonn and BonnA potentials
 and $\Lambda$=3 and 3.5$fm^{-1}$.}
\end{figure}

\begin{figure}[here]     
\scalebox{0.42}{\includegraphics[angle=-90]{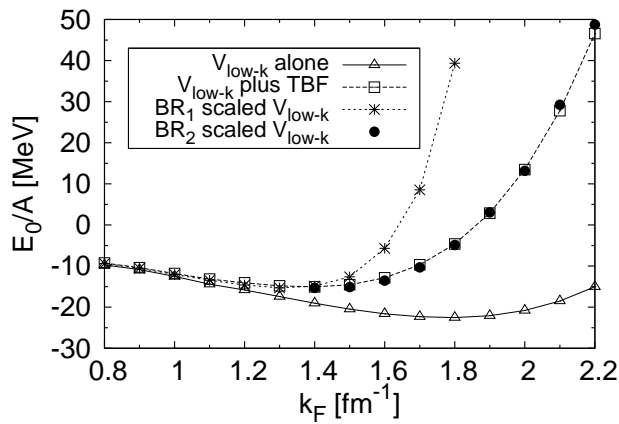}}
\caption{Ring-diagram {\rm EOS}s for symmetric nuclear matter  
given by $V_{low-k}$ alone,
 $V_{low-k}$ with linear ($BR_1$) and nonlinear ($BR_2$) scalings, 
and $V_{low-k}$ plus the  three-body
force (TBF) $V_{3b}$ of Eq.(13). 
$\Lambda$=3.5$fm^{-1}$ used for all cases. See text for other
explanations.}
\end{figure}     

  We have carried out $V_{low-k}$ ring-diagram calculations for symmetric
nuclear matter using several {\rm NN} potentials (CDBonn \cite{cdbonn}, 
Nijmegen \cite{nijmegen}, Argonne V18 \cite{argonne}
and BonnA \cite{mach89}) with several values of $\Lambda$  ranging from
3.0 to 3.5$fm^{-1}$. In Fig. 2 we present some representative results
using CDBonn and BonnA potentials, the results for other potentials
and $\Lambda$s being very similar. As shown, the results for small densities
are nearly independent of $\Lambda$ within the range considered. But for
larger densities, the results have significant variations 
with $\Lambda$ and potentials,
this being possibly a reflection of the different short-range repulsions
contained in the potential models. 
 A common feature of our results is, as 
displayed in the figure, that the calculated $E_0/A$ and saturation Fermi
momentum $k_F^0$ are all both too
high compared with the empirical values ($E_0\sim$-16 {\rm MeV} and 
$k_F^0\sim 1.35 fm^{-1}$ or $n_0\sim$0.16$fm^{-3}$).

 As discussed in section III, the above situation can be largely improved
by way of using a $V_{NN}$ with {\rm BR} scaling. In \cite{rapp99} authors
have carried {\rm DBHF} calculations for symmetric matter  using
a {\rm BR}-scaled BonnB {\rm NN} potential, and they obtained results in good
agreement with the empirical values, largely improved
over those from the unscaled potential. 
In \cite{siu09}, ring-diagram {\rm EOS} calculations for symmetric nuclear matter
have been performed using the
Nijmegen potential without and with {\rm BR} scaling, the latter giving highly
 improved results for nuclear matter saturation.
It should be useful  if the above effect on nuclear saturation
from {\rm BR} scaling also holds for other {\rm NN} potentials.
 To study this, we use  in the present work a different potential, 
the BonnA potential \cite{mach89},
for investigating the effect of {\rm BR} scaling on ring-diagram calculations for
symmetric nuclear matter. 
In Fig.3, results of such ring-diagram calculations for symmetric nuclear matter
with and without {\rm BR} scaling are presented. For the scaled calculation, 
the mesons ($\rho,\omega,\sigma$) of the BonnA potential are slightly scaled 
according to Eq.(11)
with the choice of $C_{\rho}=0.113$, $C_{\omega}=0.128$ and $C_{\sigma}=0.102$.
These values are chosen so that the calculated  $E_0/A$ and $k_F^0$ are in 
satisfactory agreement with the empirical values.  
 The {\rm EOS} given by the above {\rm BR}-scaled potential is shown by the top curve 
of Fig. 3 (labelled as `$BR_1$'),
and it has $E_0/A$=-15.3 {\rm MeV} and $k_F^0$=1.33 $fm^{-1}$, in good agreement
with the empirical values. In addition, it has compression modulus 
$\kappa$=225 {\rm MeV}. The result using $V_{low-k}$ alone is also shown
in Fig.3 (bottom curve).
Clearly {\rm BR} scaling is also important and helpful for the BonnA potential
in reproducing empirical nuclear matter saturation properties.

We shall now discuss if the above  effect of {\rm BR} scaling can be simulated by
an empirical three-body force of the Skyrme type. It is generally agreed that
the use of two-body force alone can not satisfactorily describe nuclear
saturation, certain three-body forces are needed to ensure nuclear
saturation \cite{bogner05}. 
There are basic similarities
between three-body force and {\rm BR} scaling. 
To see this, 
let us consider a meson exchanged
between two interacting nucleons. When this meson interacts with a third
spectator nucleon, this process contributes to {\rm BR} scaling or equivalently
it generates the three-body  interaction. 
In \cite{siu09}, it was already
found the ring-diagram results of {\rm BR}-scaled $V_{low-k}$ derived from
the Nijmegen potential can be well reproduced
by the same calculation except the use of the interaction given by
the sum of the $V_{low-k}$  plus 
the empirical three-body force (TBF) $V_{3b}$ of Eq.(13). 
(Note that  $V_{3b}$ is calculated using
Eq.(14) with $n$ being the local nuclear matter density.)
 Here we repeat this calculation using a different
potential, namely the BonnA potential. The strength parameter $t_3$ is adjusted
so that the low-density $(< \sim n_0)$ {\rm EOS}  given by the ($V_{low-k}$+$V_{3b}$)
 calculation are in good agreement with that from the {\rm BR}-scaled $V_{low-k}$.
(We fix the parameter $x_3$ of Eq.(14) as zero, corresponding to treating the
$^1S_0$ and $^3S_1$ channels on the same footing.)
 Results for  such a calculation, with
$t_{3}$ chosen as 2000 {\rm MeV}$fm^6$ are presented
as the middle curve of  Fig. 3 (labelled as `$V_{low-k}$ plus TBF'). 
As shown, for $k_F\leq \sim 1.4 fm^{-1}$ they
agree very well with the results from the {\rm BR}-scaled $V_{low-k}$.
The above ($V_{low-k}$+$V_{3b}$) calculation gives 
$E_0/A$=-14.7 {\rm MeV} and $k_F^0=1.40 fm^{-1}$ in satisfactory agreement 
with the {\rm BR}-scaled
results given earlier. Its compression modulus is $\kappa$=140 {\rm MeV}.

 It should be noticed, however, for $k_F$  $>\sim 1.4 fm ^{-1}$ 
the curve for `$BR_1$ scaled $V_{low-k}$' rises much more rapidly  
(more repulsive) than the `$V_{low-k}$ plus TBF' one. The compression modulus
given by them are also quite different, 225 versus  140 {\rm MeV}.
 These differences may be related to the linear {\rm BR}-scaling
adopted in Eq.(11).   This scaling is to be used for density 
less than $\sim n_0$. For density significantly larger than $n_0$,
such as in the interior of neutron star, 
this linear scaling is clearly not suitable.

To our knowledge, how to scale the mesons  at high densities is 
still an open question \cite{brown91,hatsuda,brown96,brown02,brown04}. 
In the present work, we have considered two schemes for extending
the {\rm BR} scaling to higher densities: One is  the above Skyrme-type
extrapolation; the other is an empirical  modification where in the 
high density region a nonlinear  scaling is assumed, namely  
$m^*/m=(1-C(n/n_0)^{B})$ with $B$ chosen empirically.  The  exponent $B$ 
is  1 in the linear {\rm BR} scaling of Eq.(11).
As seen in Fig. 3, the linear {\rm BR}-scaled {\rm EOS}  agrees well with
the `$V_{low-k}$ plus TBF' {\rm EOS} only in the low-density ($<\sim n_0$)
region, but not so
for densities beyond. Can a different choice of $B$ give better agreement
for the high-density region? As seen in Fig. 3, to obtain such better agreements
we need to use a scaling with weaker density dependence than $BR_1$.
Thus we have considered $B<1$, and have found that
the {\rm EOS}s with $B$ near 1/3 have much improved agreements with the Skyrme 
{\rm EOS}
in the high density region. To illutrate, we have
repeated the `$BR_1$' {\rm EOS} of Fig. 3 with only one change, namely changing 
 $B$ from 1 to 0.3. (The scaling parameters $C$s are not changed, 
for convenience of comparison.) The new results, labelled as `$BR_2$', 
are also presented in Fig.3. As seen, `$BR_2$' and `$V_{low-k}$ plus TBF' are
nearly identical in a wide range of densities beyond $\sim n_0$. This is
an interesting result, indicating that below $\sim n_0$ 
the  `$V_{low-k}$ plus TBF' {\rm EOS} corresponds to the linear $BR_1$ scaled {\rm EOS}
while beyond $\sim n_0$ the nonlinear $BR_2$ one.
The $BR_1$ and $BR_2$ {\rm EOS}s have a small discontinuity (in slope) at $n_0$,
and the above {\rm EOS} with TBF is practically a continuous {\rm EOS} with good
 fitting to both. As we shall discuss later, the above three-body force
is also important and desirable for neutron-star calculations
involving much higher densities. Possible microscopic connections
between the Skyrme three-body force and {\rm BR} scalings are being further studied,
and we hope to report our results soon in a separate publication.

 The ring-diagram nuclear matter
 {\rm EOS}s using the `$V_{low-k}$ plus TBF' interaction are in fact 
rather insensitive to the choice $\Lambda$.
 As discussed earlier,  a suitable range for $\Lambda$ is from $\sim3$
to $\sim 3.5 fm$. So in carrying out the above calculations, one first chooses
a $\Lambda$ within the above range. Then $t_3$ is determined by the requirement
that the low-density ($<\sim n_0$) ring-diagram {\rm EOS} given by $BR_1$-scaled
$V_{low-k}$ is reproduced by that from ($V_{low-k}$+$V_{3b}$).
In Fig. 4 we present some sample results  for $\Lambda$= 3 and 3.5
$fm^{-1}$ with CDBonn and BonnA potentials, all using $t_3$=2000 {\rm MeV}$fm^6$. 
Note that this $t_3$ value is for $\Lambda$=3.5 $fm^{-1}$ and BonnA 
potential; for convenience in comparison it is here 
used also for the other three cases.
It is encouraging to see that within the
above $\Lambda$ range our results are remarkably stable with 
regard to the choice
of both $\Lambda$ and $t_3$. The four curves of Fig. 4 are nearly
overlapping, and their $(E_0/A,k_F^0,\kappa)$ values are all
 close to (-15 {\rm MeV}, 1.40$fm^{-1}$, 150 {\rm MeV}). We have repeated the above
calculations for the Nijmegen and Argonne V18 potentials, and have obtained
highly similar results. As we shall discuss in the next
subsection, the inclusion of $V_{3b}$ is also important in giving a
 satisfactory neutron-matter ring-diagram {\rm EOS}. Calculations of neutron
star properties using
  the above ($V_{low-k}$ + $V_{3b}$)
interaction will also be presented there. 
Unless otherwise specified,
we shall use from now on $\Lambda$=3.5 $fm^{-1}$ for the decimation scale
and $t_3$=2000 {\rm MeV}$fm^6$ for the three-body force $V_{3b}$.  

\begin{figure}[here]     
\scalebox{0.42}{\includegraphics[angle=-90]{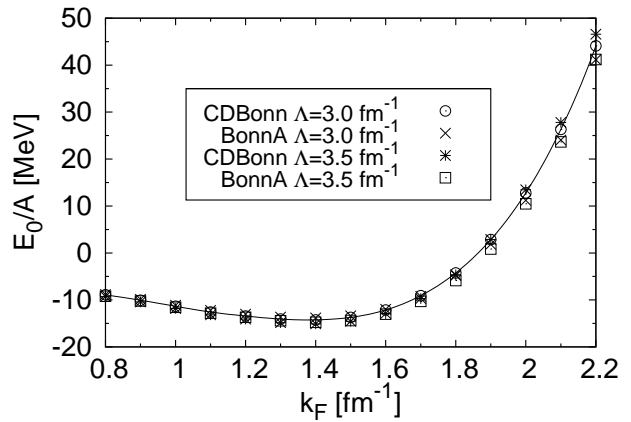}}
\caption{Ring-diagram {\rm EOS} for symmetric nuclear matter with the interaction
being the sum of  $V_{low-k}$  and the three-body force of Eq.(13). 
Four sets of results are shown for CDBonn and BonnA potentials 
with $\Lambda$=3 and
 3.5 $fm^{-1}$. A common three-body force of $t_3$=2000 {\rm MeV}$fm^6$ is 
employed.}
\end{figure}          
\subsection{Neutron star with neutrons only}
As a preliminary test of our ring-diagram {\rm EOS}, in this subsection 
we shall  consider neutron 
stars as composed of pure neutron matter only. This simplified
structure is convenient for us to describe our methods of calculation.
In addition, this also enable us to check how well can the properties of 
neutron stars be described under the pure-neutron matter assumption.
Realistic neutron stars have of course more complicated compositions; 
they  have nuclei crust and their interior
 composed of neutrons as well as other elementary particles
\cite{latti07,heisel00}. 
We shall study the effects of using $\beta$-stable  and nuclei-crust {\rm EOS}s  
in our neutron star calculations in the next subsection. 
In the present work we consider neutron stars at zero
temperature.  

Using the methods outlined in section II, we first calculate the 
ground-state energy per particle $E_0/A$ for neutron matter. Then
the energy density $\varepsilon$, inclusive of the rest-mass energy, is obtained
as 
\begin{equation}
\varepsilon(n)=n(\frac{E_0}{A}+m_nc^2)
\end{equation}
where $c$ is the speed of light and $m_n$ the nucleon mass. 
By differentiating $E_0/A$ with density, 
we obtain the pressure-density relation
\begin{equation}
p(n)=n^2\frac{d(E_0/A)}{dn}.
\end{equation}
From the above two results, the {\rm EOS} $\varepsilon (p)$ is obtained. It is the
{\rm EOS} $\varepsilon (p)$ which is used in the solution of the {\rm TOV} equations.

  To accommodate the high densities in the interior of neutron stars,
we have chosen $\Lambda$=3.5 $fm^{-1}$ for our present neutron
star calculation. Our ring-diagram {\rm EOS} for neutron matter is then calculated
using the interaction ($V_{low-k}$ +$V_{3b}$) with the parameter $t_{3}$=
2000 {\rm MeV}$fm^6$. Note this value was determined for symmetric nuclear matter,
as discussed in subsection A. Is this $t_3$ also appropriate for
the neutron matter {\rm EOS}? We shall address this question here.
 In Fig. 5 we present
results from the above neutron matter {\rm EOS} calculations for four
interactions (CDBonn, Nijmegen, Argonne V18, BonnA). It is seen that the {\rm EOS}s
given by them are quite close to each other, giving a nearly unique
neutron-matter {\rm EOS}. Friedman and Pandharipande (FP) 
\cite{pandar} have carried out variational many-body calculations 
for neutron matter {\rm EOS} 
using the two- and three-nucleon interactions $V_{14}$ and TNI respectively, 
 their {\rm EOS} results
 also shown in Fig. 5.  Brown \cite{babrown} has carried out extensive
studies of neutron matter {\rm EOS}, and has found that the FP {\rm EOS} can be reproduced
by the {\rm EOS} given by certain empirical Skyrme effective interactions 
(with both two- and three-body parts).
As seen in Fig 5,  our results agree with the FP {\rm EOS}
impressively well. 
 For comparison, we present
in Fig. 5 also the CDBonn {\rm EOS} without the inclusion of $V_{3b}$ (i.e.
$t_3$=0). It is represented by the dotted-line, and is much lower
than the FP {\rm EOS}, particularly at high densities. For $n<\sim n_0/2$
the effect of $V_{3b}$ is rather small, and in this density range
one may calculate the {\rm EOS} using $V_{low-k}$ alone.
Clearly the inclusion of $V_{3b}$ with $t_3$=2000 {\rm MeV}$fm^6$ is  
essential for attaining the above good agreement between our {\rm EOS}s
and the FP one.  It is of interest that the $t_3$
 value determined for symmetric nuclear matter turns out to be also appropriate
for neutron matter.

 In Fig. 6, our results for the {\rm EOS} $\varepsilon (p)$ are presented,
where the {\rm EOS}s given by various potentials are remarkably close to each other. 
 The inclusion of $V_{3b}$ is found to be also important here. As also
shown in Fig. 6, the {\rm EOS} given by $V_{low-k}$ alone (without $V_{3b}$)
lies considerably higher than those with $V_{3b}$. It is of interest
that for a given pressure, the inclusion of $V_{3b}$ has large effect
in reducing the energy density. 
We have chosen to use $\Lambda$= 3.5 $fm^{-1}$, and
this limits the highest pressure $p_{\Lambda}$ 
which can be provided by our ring-diagram
{\rm EOS} calculation. As shown in the figure,
the highest pressure there is about 650 {\rm MeV}/$fm^{3}$. But in neutron star
calculation we need {\rm EOS} at higher pressure such as 1000 {\rm MeV}/$fm^{3}$
(or $\sim 4\times 10^{-4} M_{\odot}c^2/km^3$). The {\rm EOS} at such high pressure
is indeed uncertain, and some model {\rm EOS} has to be employed. In the present
work we shall adopt a polytrope approach, namely we fit  a 
section of the calculated {\rm EOS} near the maximum-pressure end
by a polytrope $\varepsilon(p)=\alpha p^{\gamma}$ and use this polytrope
to determine the energy density for pressure beyond $p_{\Lambda}$.
(In our fitting  the section is chosen as $(\sim 0.8~to~1)p_{\Lambda}$.)
The polytrope {\rm EOS} has been widely and successfully used in neutron 
star calculations \cite{silbar,cooper}. In fact we have found that our
calculated {\rm EOS}, especially the section near its high-pressure end,
 can be very accurately
fitted by a polytrope. In Table I we list the polytropes obtained from
the above fitting for four {\rm NN} interactions. It is seen that the four polytropes
are close to each other. The exponent $\gamma$
plays an important role in determining the neutron-star maximum mass.

\begin{figure}[h]     
\scalebox{0.42}{\includegraphics[angle=-90]{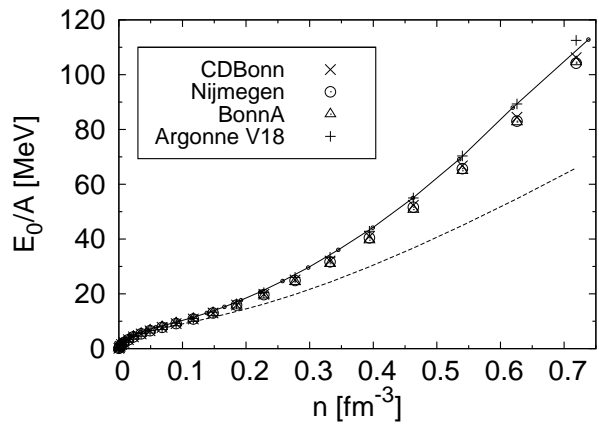}}
\caption{Ring-diagram neutron matter {\rm EOS} obtained from four realistic 
{\rm NN} potentials. The interaction `$V_{low-k}$  plus TBF' is used.
The solid line with filled small circles represents  
the results from the variational many-body calculation of 
Friedman-Pandharipande \cite{pandar}. The dotted line denotes the
{\rm EOS} using CDBonn-$V_{low-k}$ only (TBF suppressed).}
\end{figure}

\begin{figure}     
\scalebox{0.42}{\includegraphics[angle=-90]{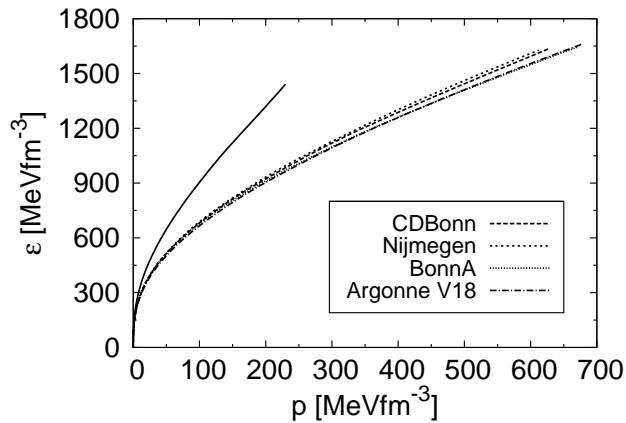}}
\caption{Neutron matter $\varepsilon(p)$ obtained from four realistic
{\rm NN} potentials. The upper-left thin line denotes the $\varepsilon(p)$
from CDBonn-$V_{low-k}$ only (TBF suppressed).}
\end{figure}

\begin{table}[ht]
\caption{Fitted polytrope $\alpha p^{\gamma}$ for high pressure region.
See text for other explanations.}
\centering
\begin{tabular}{c c c}
\hline\hline
Potentials & $\alpha^{\ast}$ & $\gamma$\\
\hline
CDBonn    &69.69 $\pm$ 1.01 & 0.4876 $\pm$ 0.0022\\
Nijmegen   & 69.99 $\pm$1.01 & 0.4885 $\pm$ 0.0021 \\
BonnA      & 72.30 $\pm$ 1.01 & 0.4779 $\pm$ 0.0021 \\
Argonne V18  & 67.71 $\pm$ 1.01 & 0.4887 $\pm$ 0.0021 \\
\hline
$\ast$ unit of $\alpha$ is $(MeV/fm^3)^{1-\gamma}.$
\end{tabular}
\label{table:fitted eos}
\end{table}

 In obtaining the neutron star properties, we numerically solve  
the {\rm TOV} equations (1) by successive integrations. In so doing, we need 
to have the pressure $P_c$ at the center of the neutron star 
to begin the integration. As we shall see soon, different
$P_c$s will give, e.g. different masses for neutron stars. We also need
the {\rm EOS} $\varepsilon (p)$ for a wide range of pressure. As discussed above,
we shall use the ring-diagram {\rm EOS} for pressure less than $p_{\Lambda}$
and the fitted polytrope {\rm EOS} for larger pressure. In Table II, we list
some typical results for  the neutron star mass $M$ and its 
corresponding 
 radius $R$ and static moment of inertia $I$. (The calculation of $I$
will be discussed later.) They were obtained with
four different center pressures $P_c$, and as seen these properties of the
neutron star vary significantly with $P_c$.

\begin{table}[ht]
\caption{Neutron stars with different center pressures.}
\centering
\begin{tabular}{c c c c}
\hline\hline
$P_c[M_{\odot}c^2/km^3]$ & $M[M_{\odot}]$ & $R[km]$ & $I[M_{\odot}km^2]$\\
\hline
$8.07 \times 10^{-7}$    & 0.101 & 11.58 & 3.78\\
$7.18 \times 10^{-6}$  & 0.347 & 10.12 & 13.51 \\
$5.38 \times 10^{-5}$    & 1.037 & 10.10 & 50.02\\
$2.33 \times 10^{-4}$   &1.597 & 10.00 & 70.69 \\
\hline
\end{tabular}
\label{table:fitted eos}
\end{table}

\begin{figure}[here]     
\scalebox{0.42}{\includegraphics[angle=-90]{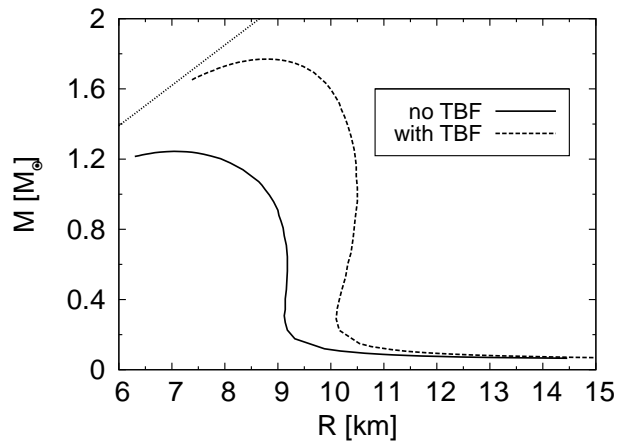}}
\caption{Mass-radius trajectories of pure neutron stars from ring-diagram 
{\rm EOS}s given by the CD-Bonn $V_{low-k}$ interaction  with and without 
the three-body force (TBF) $V_{3b}$. Only stars to the right of maximum 
mass are stable against gravitational collapse. 
Causality limit is 
indicated by the straight line in the upper left corner.}
\end{figure}
\begin{figure}[here]     
\scalebox{0.42}{\includegraphics[angle=-90]{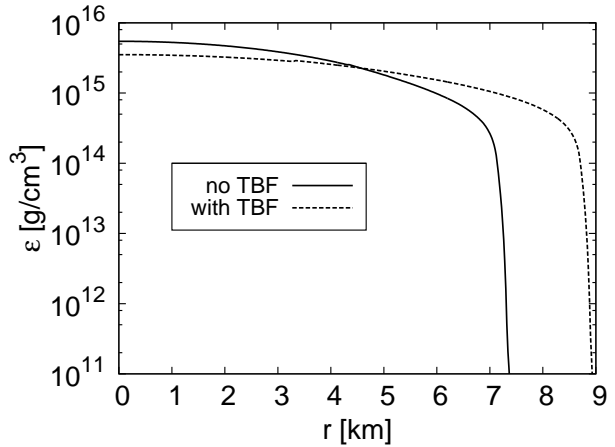}}
\caption{Density profiles of maximum-mass neutron stars of Fig. 7.}
\end{figure}
 
We present some of our calculated results for the 
mass-radius trajectories of neutron stars in Fig. 7. They were obtained
using the CDBonn $V_{low-k}$ ($\Lambda$=3.5 $fm^{-1}$) with and without
the three-body force $V_{3b}$ ($t_3$=2000 {\rm MeV}$fm^6$) discussed earlier. As seen,
the inclusion of $V_{3b}$ significantly increases both the maximum neutron
star mass $M$ and its corresponding radius $R$; the former increased
from $\sim 1.2$ to $\sim 1.8M_{\odot}$ and the latter from $\sim 7$ to
$\sim 9$km. The above results are understandable, because $V_{3b}$ makes the
{\rm EOS} stiffer and consequently enhances both $M$ and $R$. Note that our results
are within the causality limit.
 We have repeated the above calculations using the Nijmegen, Argonne
and BonnA potentials, with results quite similar to the CDBonn ones.
 In Fig. 8, we present the density profiles corresponding to the 
maximum-mass neutron stars of Fig. 7. It is clearly seen that the inclusion
of the three-body force TBF
has important effect in neutron-star's density distribution, reducing
the central density and enhancing the outer one. 

We have also performed calculations using the $BR_1$-scaled 
$V_{low-k}$ interaction (BonnA and $\Lambda$=3.5 $fm^{-1}$)
without $V_{3b}$. The resulting maximum mass and its radius
given  are respectively $\sim 3.2M_{\odot}$ and $\sim 12$km, 
both being considerably larger than the values of Fig. 7. This is also
reasonable, because, as was shown in
Fig. 3 the $BR_1$-scaled {\rm EOS} is much stiffer than the `$V_{low-k}$ plus TBF'
one. It may be mentioned that if the neutron-matter
{\rm EOS} given by the {\rm BR}-scaled interaction is plotted in Fig. 5, it would be
very much higher, especially in the high
density region, than the FP {\rm EOS} shown there. However,
the `$V_{low-k}$ plus TBF' ones are very close to the FP one as shown
earlier. We feel that the above comparison is a further indication that
the linear $BR_1$ scaling of Eq.(11) is not suitable for high density.
It is suitable only for density up to about $\sim n_0$. 

     Moment of inertia  is an  important property of
 neutron stars.\cite{schutz,worley} 
Here we would like to calculate this quantity using our $V_{low-k}$
ring diagram formalism. Recall that we have used the {\rm TOV} equations (1) 
to calculate neutron star mass and radius, and in so doing we also obtain
the density distribution inside neutron stars. From this distribution,
 the moment of inertia $I$ of neutron stars is readily calculated. 
It may be noted that 
the {\rm TOV} equations are for spherical and 
static (non-rotating) neutron stars,
and the $I$ so  obtained is the static one for spherical neutron stars.
The moment of inertia for rotating stars are more complicated to
calculate, but  for 
low rotational frequencies (less than $\sim 300$Hz) they are rather 
close to the static ones \cite{worley}. In Fig. 8, we present our
 results for two calculations, the interactions used being the same
as in Fig. 7. It is seen that the the inclusion of our three-body force
$V_{3b}$ (TBF) largely enhances the moment of inertia of maximum-mass
neutron star.

  The measurement of neutron-star moment of inertia is still
 rather uncertain, and the best determined value so far is that of the Crab 
pulsar (97$\pm$ 38 $M_{\odot} km^2$) \cite{bejger}. 
For $M \geq 1.0 M_{\odot}$, Lattimer and Schutz \cite{schutz} have determined
an empirical formula relating the moment of inertia $I$ of 
neutron stars to their
mass $M$ and radius $R$, namely
\begin{eqnarray}
I &\approx & (0.237 \pm 0.008)MR^2 \nonumber \\
 &&\times(1+4.2\frac{M}{M_{\odot}}\frac{km}{R}+90(\frac{M}{M_{\odot}}
\frac{km}{R})^4).
\end{eqnarray}
To check if our calculated ($M,~R,~I$) are consistent with this empirical
relation, we have
computed $I$  using our calculated $M$ and $R$ values  
(with TBF as in the top curve of Fig. 9) as inputs to Eq.(17). 
Results of this computation are also
shown in Fig. 9. As shown, they are in  good general agreement
with the empirical formula. Especially our moment of inertia at maximum mass 
agrees remarkably well with the corresponding empirical value.
 We have also repeated the above computation with
 other potentials and obtained similar results. 

\begin{figure}[h]  
\scalebox{0.42}{\includegraphics[angle=-90]{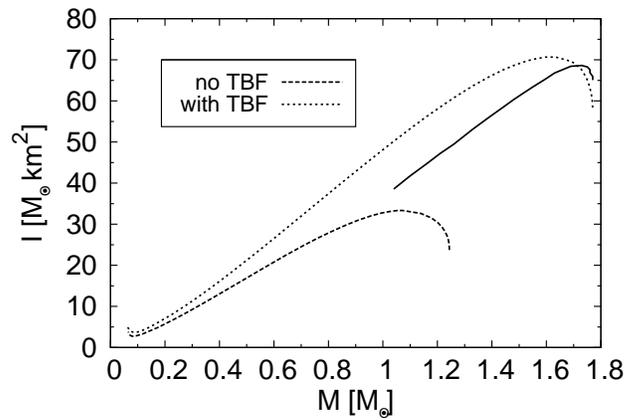}}
\caption{Pure neutron stars' moments of inertia  calculated 
 from CDBonn potential with and without three-body force (TBF).
Results from the empirical formula (17) for $M \geq 1.0 M_{\odot}$ 
with TBF are denoted by the solid line.}
\end{figure}

\subsection{Effects from $\beta$-stable and nuclei-crust {\rm EOS}s}           
      
      In the preceding subsection, we considered neutron star as composed of 
neutrons only, and we have obtained rather satisfactory results. Would the
quality of them be significantly changed when we use a more realistic
composition? 
  As a small-step improvement,  in this subsection we shall first
 carry out calculations using the ring-diagram  $\beta$-stable {\rm EOS}
  composed of neutrons, protons, electrons and muons only. The results of
them will be briefly compared with those obtained with neutrons only.
Calculations using a combination of the {\rm BPS}  {\rm EOS} \cite{baym71} 
inside the nuclei crust  and our $\beta$-stable {\rm EOS} for the interior will
 also be carried out. The crust of the neutron star is composed of
two parts, the outer and inner crust 
\cite{baym71,jxu09,steiner,ruster06,horow03}. The choice of the density
regions defining these crusts and how to match the {\rm EOS}s at the boundaries
between different regions will be discussed.

Let us first discuss  our $\beta$-stable {\rm EOS}, where the composition
fractions of its constituents are determined by 
the chemical equilibrium equations 
\begin{eqnarray}
   \mu_n &=& \mu_p+\mu_e  \\
   \mu_e &=& \mu_{\mu}      
\end{eqnarray}     
together with the charge and mass conservation conditions
\begin{eqnarray}
n_p&=&n_e+n_{\mu}\\
n&=&n_n+n_p+n_e+n_{\mu}.
\end{eqnarray}
In the above, $\mu_n, \mu_p, \mu_e$ and $\mu_{\mu}$ are the chemical potentials 
for neutron, proton, electron and muon respectively, and their densities
 are  respectively  $n_n$, $n_p$, $n_e$ and $n_{\mu}$.
The total density is $n$. For a given $n$, the composition fractions
of these constituents are determined by the above equations.
Note that these equations are
solved self-consistently, since the chemical potentials and densities
are inter-dependent. We have used iteration methods for this solution.
Clearly the composition fractions are not uniform inside the 
$\beta$-stable neutron star;
they depend on the local density. For example, the proton fractions
$\chi\equiv n_p/n$ in different density regions of $\beta$-stable neutron
star are generally different. In solving the above equations, we have 
calculated the chemical potentials $\mu_n$ and $\mu_p$ using the HF
approximations as indicated by Eq.(8). Since the interactions involving
 electrons and muons are much weaker than the strong nucleon ones,
we have treated them as  free Fermi gases and in this way their chemical 
potentials are readily obtained.

Results for the proton fractions $\chi$ calculated from the above
equations are displayed in Fig. 10; they were obtained using
four  $V_{low-k}$($\Lambda$=3.5 $fm^{-1}$) interactions, all  with
the three-body force $V_{3b}$($t_3$=2000 {\rm MeV}$fm^6$). 
We note that our calculated proton
fractions are all quite small, the maximum proton fraction given by BonnA
potential being  $\sim 7\%$ and that by both Argonne V18 and Nijmegen
potentials being $\sim 2\%$. Also they exhibit
a saturation behavior, being near maximum at density between $\sim$0.6 and
$\sim$0.8 $fm^{-3}$ and diminishing to near zero on both sides. Our results
suggest that $\beta$-stable neutron star has small proton admixtures
only within a intermediate layer; the neutron star's core and surface layer
are both essentially  pure neutron matter. Proton fractions in 
$\beta$-stable neutron stars are an important topic and have been
extensively studied \cite{li08,poland}. They are closely related to the
density dependence of symmetry energy, which is being determined
in several laboratories \cite{li08}. There have been a number of
 calculations for these fractions using different many-body methods
and different interactions; their results are, however, widely different 
from each other (see Fig.1 of \cite{poland}), some of them being close to
ours. Further studies of the proton fractions would
certainly be in need and of interest.

\begin{figure}[here]    
\scalebox{0.42}{\includegraphics[angle=-90]{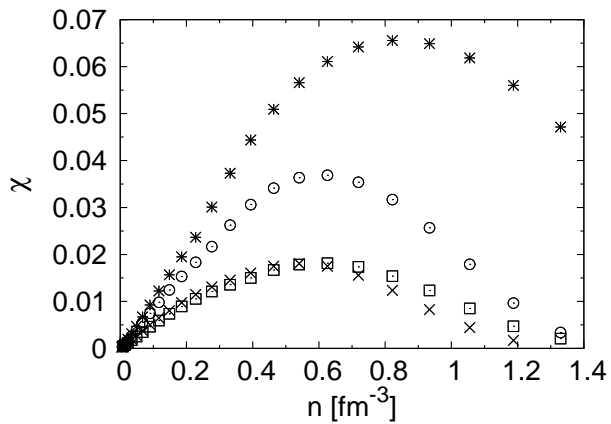}}
\caption{Proton fraction of $\beta$-stable neutron 
star from realistic {\rm NN} potentials. Symbols are 
BonnA($\ast$), CDBonn($\circ$), Argonne V18 ($\Box$) and Nijemgen  ($\times$).
The interaction `$V_{low-k}$ plus TBF' is used.}
\end{figure}

 With our calculated proton fractions, we proceed to calculate the properties
of $\beta$-stable neutron stars whose energy-density {\rm EOS} is 
$\varepsilon (n)= \varepsilon _{np} +\varepsilon_{e}+\varepsilon_{\mu}$,
where $n$ is the total density of Eq.(21).
Electrons and muons are treated as free Fermi gas and their energy densities
are readily obtained.  $\varepsilon _{np}$ is the neutron-proton energy density
to be evaluated using the ring-diagram method described in section II.
This energy density is in fact $\varepsilon_{np}(n_{np},\alpha)$
where $n_{np}=n_n+n_p$, the combined nucleon density, and $\alpha$ is the
asymmetric parameter of Eq.(5). Calculations for $\beta$-stable neutron
stars are more complicated than the pure neutron matter case of subsection
B, for which $\alpha=1$ independent of the total density $n$. In contrast,
here we need to calculate $\varepsilon_{np}$
 for many $(n_{np},\alpha)$ values since they are dependent on $n$ (see
Eqs.(18-21)). Then the {\rm EOS}  $\varepsilon_{np}(p)$, which expresses 
energy density in terms of pressure $p$, is obtained by density
differentiations of $\varepsilon_{np}(n_{np},\alpha)$, similar 
to what we did in subsection A. Then by solving the {\rm TOV} equations,
the various properties of $\beta$-stable neutron stars are obtained.

\begin{figure}[here]   
\scalebox{0.42}{\includegraphics[angle=-90]{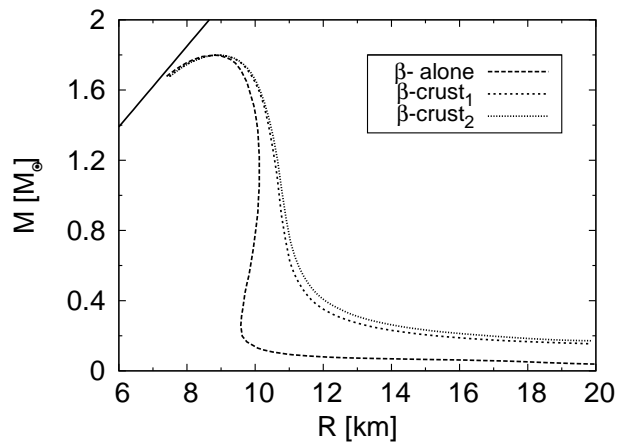}}
\caption{Mass-radius trajectories of neutrons stars
obtained using only the ring-diagram $\beta$-stable {\rm EOS}  ($\beta$-alone)
and  a combination of the ring and nuclei-crust {\rm EOS}s with
inner-crust boundary $n_t$= 0.04 ($\beta$-crust$_1$)
and 0.05 $fm^{-3}$  ($\beta$-crust$_2$).
 The `CDBonn-$V_{low-k}$ plus TBF' potential is used for the ring {\rm EOS}.
See the caption of Fig. 7 for other explanations. }
\end{figure}

 To illustrate our results for the mass-radius trajectories  for 
the $\beta$-stable
neutron stars consisted of $(n,p,e,\mu)$, we have performed such a
calculation  for the CDBonn
potential and present its results in Fig. 11 (labelled `$\beta$-alone'). 
As seen it is quite similar to the corresponding one of Fig. 7 
for the pure neutron {\rm EOS}. The trajectories using the same method but with
other {\rm NN} potentials (Nijmegen, Argonne and BonnA) have also been 
calculated, and are also very similar to the corresponding pure-neutron
ones.
This close similarity indicates that the effect 
from the admixture of $(p,e,\mu)$ is not important as based on our present
calculation,  this being largely due to the smallness
of the proton fractions discussed earlier.  

\begin{figure}[here]   
\scalebox{0.42}{\includegraphics[angle=-90]{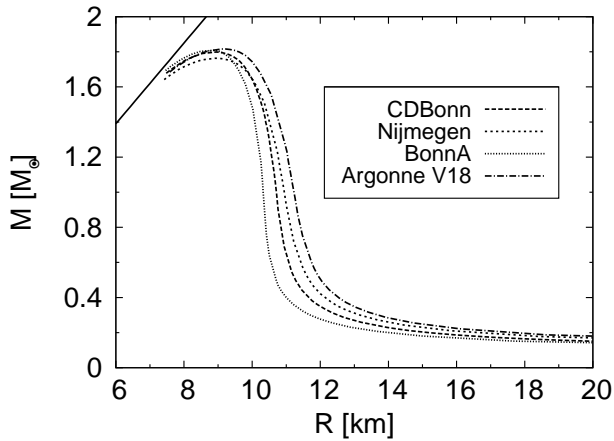}}
\caption{Mass-radius trajectories of neutron stars 
calculated with a combination of $\beta$-stable ring-diagram {\rm EOS} for the core
and the nuclei-crust {\rm EOS}s for the crusts.
 Ring-diagram {\rm EOS}s given by four {\rm NN} 
potentials, all with 
the $V_{3b}$ three-body force, 
$\Lambda$=3.5$fm^{-1}$ and $t_3$=2000 {\rm MeV}$fm^6$ are used. 
$n_t$=0.04 $fm^{-3}$ is used for the inner crust boundary.
See the caption of Fig. 7 for other explanations.
}
\end{figure}
\begin{figure}[th]   
\scalebox{0.42}{\includegraphics[angle=-90]{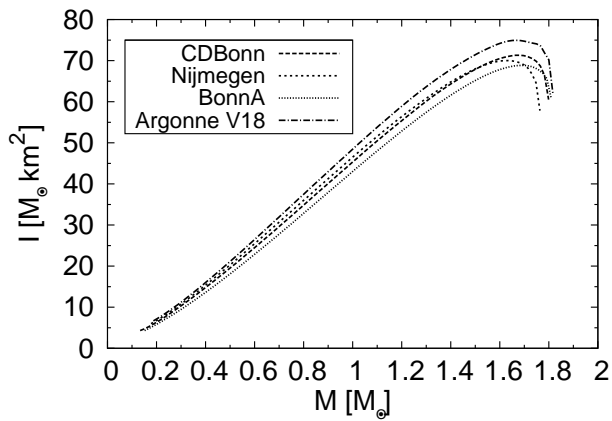}}
\caption{Moments of inertia of neutron stars of Fig. 12.
}
\end{figure}

\begin{table}[here]
\caption{Maximum mass and the corresponding radius and moment of inertia of  
$\beta$-stable neutron stars with nuclei-crust boundary $n_t$=0.04 $fm^{-3}$.
The three-body force $V_{3b}$ is included for the results in the first
four rows, but is not in the last.}
\centering
\begin{tabular}{c c c c}
\hline\hline
Potentials & $M[M_{\odot}]$ & $R[km]$ & $I[M_{\odot}km^2]$\\
\hline
CDBonn    & 1.80 & 8.94 & 60.51\\
Nijmegen & 1.76 & 8.92 & 57.84 \\
BonnA    & 1.81 & 8.86 & 61.09\\
Argonne V18  &1.82 & 9.10 & 62.10 \\
\hline
CDBonn($V_{3b}$=0)    & 1.24 & 7.26 & 24.30\\
\hline
\end{tabular}
\label{table:fitted eos}
\end{table}

 So far we have carried out microscopic calculations of neutron stars
with the assumption  that they are 
made of a homogeneous medium 
composed of neutrons, protons, electrons and muons. It is believed that
the  crust
of neutron stars is not of such homogeneous medium; it is "nuclei crust" 
where nucleons are clustering into nuclei \cite{baym71,jxu09}.
Here we would like to make some estimates on the nuclei-crust corrections
to our present calculations. In our estimates, we employ three different
{\rm EOS}s for the outer-, inner-crust and core regions. These regions 
refer to the density regions  $n<n_{out}$, $n_{out}<n<n_{t}$
and $n>n_{t}$ respectively, with $n_{out}=2.57\times 10^{-4}fm^{-3}$
\cite{ruster06,jxu09}. $n_{t}$ is the transition density
separating the  inner crust and the homogeneous core, and  several
models have been employed to determine its value \cite{jxu09,steiner}.
For the outer-crust region  we use the well-known  
{\rm BPS} nuclei-{\rm EOS} \cite{baym71}. 
For the core region, our $\beta$-stable ring-diagram {\rm EOS} will be employed.
The {\rm EOS} in the inner-crust region is somewhat uncertain, and so is the
transition density separating the inner crust and core.
 We shall use in our calculations $n_{t}=0.05$ and $0.04fm^{-3}$
\cite{jxu09,steiner}
to illustrate the effect of the nuclei crust. Following
 \cite{horow03,jxu09}, we use in the inner-crust region a polytropic 
{\rm EOS}, namely
$p=a+b\varepsilon^{4/3}$ with the constants $a$ and $b$ determined
by requiring a continuous matching of the three {\rm EOS}s at $n_{out}$ and
$n_{t}$.

 In Fig. 11, our results for the mass-radius trajectories using the
above three {\rm EOS}s with $n_t$=0.04 and 0.05 $fm^{-3}$, labelled
$\beta$-crust$_1$ and $\beta$-crust$_2$ respectfully, are compared 
with the trajectory
given by the $\beta$-stable alone (namely $n_t$=0).
As seen, the effect from the nuclei-crust {\rm EOS} on the maximum neutron-star 
mass and its radius  is rather small, merely increasing  the maximum mass 
by $\sim 0.02M_{\odot}$ and its radius  $\sim 0.1$km 
as compared with the $\beta$-alone ones. 
However, its effect is important in the low-mass large-radius
region, significantly enhancing the neutron-star mass there.
That the maximum mass is not significantly changed by the inclusion
nuclei-crust {\rm EOS} is consistent with  Fig. 8 which indicates 
the mass of maximum-mass neutron stars being confined predominantly
in the core region.
It may be mentioned that our ring-diagram {\rm EOS} is microscopically
calculated from realistic {\rm NN} interactions, while the crust {\rm EOS}s are not.
So there are disparities between them. It would be useful and 
of much interest if the crust {\rm EOS}s can also be derived from realistic
{\rm NN} interactions using similar microscopic methods. Further studies
in this direction are needed. 
 
   In. Fig.12 we present our mass-radius results using the above three
{\rm EOS}s with $n_t$=0.04 $fm^{-3}$. Four {\rm NN} potentials are employed, and
they give similar trajectories, especially in the high- and low-mass regions.
 A corresponding comparison for the moment of inertia
is presented in Fig. 13; again the results from the four potentials
are similar.
In Table III, our results for
the maximum neutron mass and its radius and moment of inertia
using the above combined {\rm EOS}s 
are presented, and as seen the results for the maximum-mass neutron star
given by  the four potentials
are indeed close to each other. It is also seen that the effect of
the three-body force is quite important for $M,~R$ and $I$, as illustrated
by the CDBonn case.

\section{Summary and conclusion}
  We have performed neutron-star calculations based on three types
of {\rm EOS}s: the pure-neutron ring-diagram, the $\beta$-stable $(n,p,e,\mu)$
ring-diagram and the {\rm BPS} nuclei-crust {\rm EOS}.
The ring-diagram {\rm EOS}s, where the $pphh$ ring diagrams are summed 
to all orders,
are microscopically derived using the low-momentum interaction
  $V_{low-k}$  obtained
from four realistic {\rm NN} potentials (CDbonn, Nijmegen, Argonne V18, BonnA). 
We require that the {\rm EOS}
used for neutron stars should give satisfactory saturation properties
for symmetric nuclear matter, but this requirement is not met by our
 calculations using the above potentials as they are.
Satisfactory nuclear matter saturation
properties can be attained by using the above potentials
with the commonly used linear {\rm BR} scaling ($BR_1$)
where the the masses of in-medium mesons are slightly
suppressed compared with their masses in vacuum. But this linear scaling
is not suitable for neutron stars; the maximum mass of neutron star
given by our $BR_1$ ring-diagram calculation is $\sim 3.2 M_{\odot}$
which is not satisfactory. $BR_1$ is suitable only for low densities;
 it needs some extension
so that it can be applied to the high densities inside the neutron star. 
We have used an extrapolation method for this extension, namely
we add an empirical Skyrme-type
three-body force $V_{3b}$ to $V_{low-k}$. We have found that the {\rm EOS} given
 by this extrapolation agrees well with the {\rm EOS} obtained from linear $BR_1$
scaling for low densities, but for high densities it agrees well with that
 from  a nonlinear $BR_2$ scaling. The {\rm EOS} using the above extrapolation
gives satisfactory saturation properties for symmetric nuclear matter,
and for neutron matter it agrees well with the FP {\rm EOS} for neutron
matter.

 The effects from $V_{3b}$ have been
found to be both important and desirable. Compared with the results given
by the unscaled $V_{low-k}$, it increases the maximum mass of the neutron
star and its radius and moment of inertia by  $\sim 40\%$
, $\sim20\%$ and $\sim 150\%$ respectively. 
The proton fractions are found to be generally 
small ($<7\%$), making our neutron-star results using the pure-neutron {\rm EOS}
and those using the $\beta$-stable {\rm EOS} being nearly the same.
 We have estimated the effect from the  nuclei-crust {\rm EOS}s by using a combination
of three {\rm EOS}s: the {\rm BPS} {\rm EOS} for the outer crust, a fitted polytropic {\rm EOS}
for the inner crust and our $\beta$-stable ring-diagram {\rm EOS} for the core region.
The effect from the nuclei-crust {\rm EOS}s on the maximum neutron-star 
mass and its radius  is found to be rather small, as compared with those
given by the calculation where the $\beta$-stable {\rm EOS} is used throughout.
However, its effect is important in the low-mass large-radius
region, significantly enhancing the neutron-star mass there.
Using the above combined three {\rm EOS}s, our results for
neutron star's maximum mass and its radius and moment of inertia are, 
respectively, 
 $\sim 1.8 M_{\odot}$,  $\sim 9 km$ and  
$\sim 60 M_{\odot} km^2$, 
all in good agreement with accepted values.

How to extend the {\rm BR} scaling to high densities is still an open question.
Although we  have obtained satisfactory results by using a nonlinear
scaling for the high-density region, or equivalently a Skyrme-type 
three-body force, for the extension, 
it would still be certainly  useful and interesting 
to explore  other ways for doing so.
Further studies in this direction would be very helpful in 
determining the medium dependent
modifications to the {\rm NN} potentials
in the high-density region.

{\bf Acknowledgement} We  thank  G.E. Brown, Edward Shuryak, Izmail
Zahed and Shu Lin for many helpful discussions. 
This work is supported in part
by the U.S. Department of Energy  under Grant No. DF-FG02-88ER40388
and the National Science Foundation under Grant No. PHY-0099444.

\end{document}